\documentclass[aps,preprint,nofootinbib,superscriptaddress]{revtex4}
\usepackage{amssymb}

\usepackage{epsfig}

\begin{document}

\date{\today}
\title{ Fusion-fission reactions with modified Woods-Saxon potential}

\author{Ning Wang}
\email{wangning@gxnu.edu.cn} \affiliation{Institut f\"{u}r
Theoretische Physik der Universit\"{a}t, D-35392 Giessen, Germany}
\affiliation{College of Physics and Electronic Engineering, Guangxi
Normal University, Guilin 541004, P. R. China}
\author{Kai Zhao}
\affiliation{China Institute of Atomic Energy, Beijing 102413, P.
R. China}

\author{Werner Scheid}
\affiliation{Institut f\"{u}r Theoretische Physik der
Universit\"{a}t, D-35392 Giessen, Germany}

\author{Xizhen Wu}
\affiliation{China Institute of Atomic Energy, Beijing 102413, P.
R. China}
\begin{abstract}

A modified Woods-Saxon potential model is proposed for a unified description of
the entrance channel fusion barrier and the fission barrier of fusion-fission reactions based on the Skyrme energy-density functional approach. The fusion excitation functions of 120 reactions have been systematically studied. The fusion (capture) cross sections are well described with the calculated potential and an empirical barrier distribution. Incorporating a statistical model (HIVAP code) for describing the decay of the compound nucleus, the evaporation residue (and fission) cross sections of 51 fusion-fission reactions have been systematically investigated. Optimal values of some key parameters of the HIVAP code are obtained based on the experimental data of these reactions. The experimental data are reasonably well reproduced by the calculated results. The upper and lower confidence limits of the systematic errors of the calculated results are given.

\end{abstract}

\maketitle


\begin{center}
\textbf{I. INTRODUCTION}
\end{center}

The production of superheavy nuclei as evaporation residues in
fusion reactions is a field of very intense studies in the recent
decades \cite{shen02,Gup05,Zag01a,Ada98,Reis92,Feng06}. So far, the
superheavy elements $Z=110\sim 116$ and 118 have been synthesized
\cite{Hof95,Hof95a,Hof00,Ogan00,Ogan00a,Ogan04,Ogan04a,Mori04a,Ogan06}.
Theoretical support for these very time consuming experiments is
vital in choosing the optimum target-projectile-energy combinations,
and for the estimation of cross sections and identification of
evaporation residues. A self-consistent microscopic dynamics model
is still not yet available for practical studies of the whole fusion
process from the capture to the decay of the heavy compound nuclei.
Therefore, in the practical calculation of the evaporation residue
cross section, the reaction process leading to the synthesis of
superheavy nuclei is divided into two or three steps. Firstly, the
projectile is captured by the target and a dinuclear system is
formed which then evolves into the compound nucleus, and finally,
the compound nucleus loses its excitation energy mainly by emission
of particles and $\gamma$-ray and goes to its ground state. The
simplified version of the evaporation residue cross section is given
by
\begin{eqnarray}
\sigma_{\rm ER}(E_{\rm c.m.}) = \sigma_{\rm cap}(E_{\rm c.m.}) P_{\rm CN}(E_{\rm c.m.}) W_{\rm sur}(E_{\rm c.m.}).
\end{eqnarray}
Here, $\sigma_{\rm cap}$, $P_{\rm CN}$ and $W_{\rm sur}$ are the
capture cross section for the transition of the colliding nuclei
over the entrance channel Coulomb barrier, the probability of the
compound nucleus formation after the capture and the survival
probability of the excited compound nucleus, respectively. There are
several unsolved questions in each component of the right side of
Eq.(1) which leave a certain margin of uncertainty in the estimates
of the evaporation residue cross section \cite{Siw05}. In addition,
there could be several  parameters in the practical models which are
hardly unambiguously determined by a very limited number of measured
evaporation residue cross sections of superheavy nuclei. For
example, the calculated formation probability $P_{\rm CN}$ of the
compound nuclei for reaction $^{58}$Fe+$^{208}$Pb in \cite{Zag01a}
is about two orders of magnitude larger than that obtained in
Ref.\cite{Ada98}, both of the models can, however, reproduce the
measured evaporation residue cross sections satisfactorily.
Therefore, it is necessary to test and determine the interaction and
parameters adopted in each component of Eq.(1) individually.

To study the three components in Eq.(1) individually, we first
investigate the influence of the fission and quasi-fission on the
fusion-fission reactions. It is generally thought that for systems
with the compound-nuclear charge number $Z_{\rm CN}$ smaller than
about 60, the fission barrier is high enough to make fission an
improbable decay mode at incident energies close to the fusion
barrier \cite{Reis85}. Thus for these reactions, $\sigma_{\rm ER}
\simeq \sigma_{\rm fus} \simeq \sigma_{\rm cap}$ holds at
near-barrier energies. To see it more clearly, we present a
schematic figure (Fig.1(a)). The horizontal and vertical axis denote
the compound-nuclear charge number $Z_{\rm CN}$ and the mass
asymmetry of the reaction system $\eta=(A_2-A_1)/(A_2+A_1)$,
respectively. Here, $A_1$ and $A_2$ are the projectile and target
masses. The fusion reactions in region I have $P_{\rm CN} \simeq
W_{\rm sur} \simeq 1$ as discussed before. There are quite a large
number of experimental data of evaporation residue cross sections
for the reactions in region I accumulated in recent decades, which
makes it possible to establish a reliable model for systematic
description of the capture process without the influence of fission
and quasi-fission. For heavier compound systems the fission
increases rapidly with the $Z_{\rm CN}^{\; 2}/A_{\rm CN}$ and the
angular momentum. For sufficiently asymmetric systems with $Z_{\rm
CN}$ well below 100 (systems in region II of Fig.1(a)), and at
energies close to the fusion barrier, it is generally recognized
that  $\sigma_{\rm fus}=\sigma_{\rm ER}+\sigma_{\rm FF}$. Here the
$\sigma_{\rm fus}$, $\sigma_{\rm ER}$ and $\sigma_{\rm FF}$ are the
cross sections for fusion, evaporation residue and fission,
respectively. For systems in region II, it is thought that the
quasi-fission barrier is high enough and thus $P_{\rm CN} \simeq 1$.
The available experimental data of the evaporation residue cross
sections for reactions in region II are less than those in region I,
but they seem to be much enough for a systematic investigation to
test and determine some key parameters of a statistical model for
calculating the survival probability $W_{\rm sur}$, combining the
model for describing the capture cross sections, without the
influence of the quasi-fission. In addition, the measured fusion
cross sections for reactions in region II can further test the
theoretical model for calculating $\sigma_{\rm cap}$.  For $Z_{\rm
CN}$ larger than about 100 (systems in region III of Fig.1(a)),
quasi-fission occurs. Thus in the calculation of the evaporation
residue cross sections for these reactions, the influence of
quasi-fission should be taken into account ($P_{\rm CN}<1$). In
Fig.1(b) we illustrate this point more clearly. We show the contour
plot of the quasi-fission barrier heights of the reactions with
reaction parters along the $\beta$-stability line. Here, the height
of the quasi-fission barrier is empirically estimated by the depth
of the pocket of the entrance channel capture potential obtained
with a modified Woods-Saxon potential which will be discussed in
Sect.II. The height of the quasi-fission barrier decreases rapidly
with the increase of the compound-nuclear charge, especially for
symmetric target-projectile combination. For reactions with the same
$Z_{\rm CN}$, those with more asymmetric target-projectile
combinations have higher quasi-fission barriers. It is expected that
for sufficiently asymmetric systems the fusion probability $P_{\rm
CN}$ is approximately equal to one as mentioned above.  If both
$\sigma_{\rm cap}$ and $W_{\rm sur}$ can be predicted reliably, this
would help to understand the dynamics of fusion and quasi-fission.

\begin{figure}
\includegraphics[angle=0,width=1.0\textwidth]{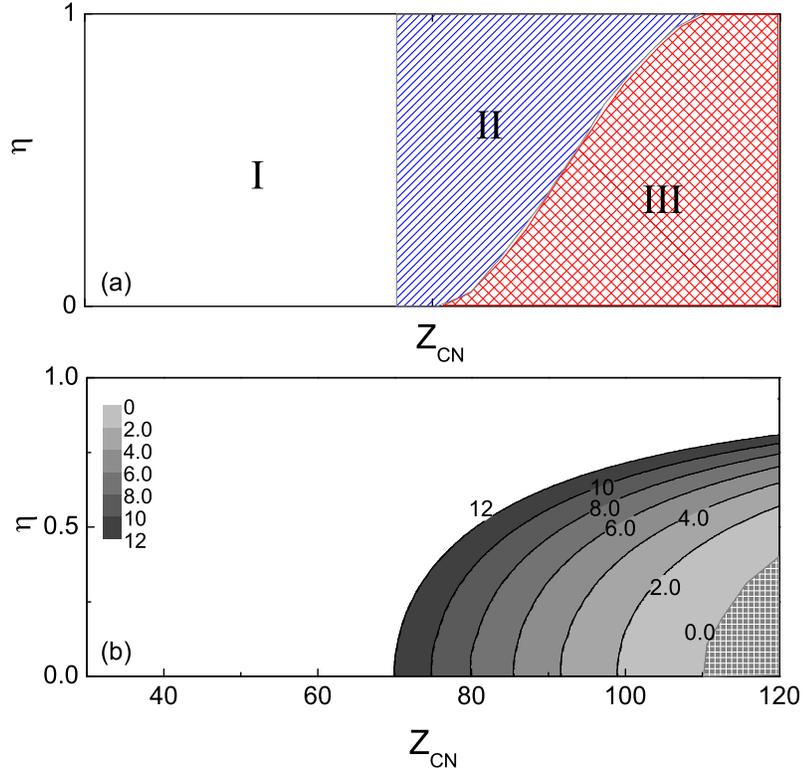}
 \caption{(Color online) (a) A schematic figure for different types of fusion reactions. The horizontal and vertical axis denote the compound-nuclear charge number $Z_{\rm CN}$ and the mass asymmetry of the reaction system $\eta=(A_2-A_1)/(A_2+A_1)$, respectively. (b) Contour plot of the quasi-fission barrier heights obtained with a modified Woods-Saxon potential (which will be introduced in the next section) for reactions with nuclei along the $\beta$-stability line.}
\end{figure}

Based on above discussions, the emphasis of this paper is put on
study of such fusion-fission reactions in which the quasi-fission is
not important. To study this kind of reactions we employ a modified
Woods-Saxon potential model based on the Skyrme energy density
functional together with the extended Thomas-Fermi approach. This
model was first proposed in \cite{Liu06} and a large number of
fusion reactions have been described satisfactorily with the
entrance channel potential. The potential between nuclei around the
touching point can be accurately evaluated with numerical algorithm
\cite{Liu06}. Unfortunately, it is not so convenient for any
practical application because one needs to evaluate numerically the
microscopic densities of the interacting nuclei, the derivatives of
these densities and the integrals. It is better to find an
analytical expression for the potential. In this work we will
present an analytical modified Woods-Saxon (MWS) form for the
potential based on the numerical results. With the analytical MWS
potential, both the fusion barrier and the fission barrier of a
reaction system will be consistently studied. For calculation of
$W_{\rm sur}$, the well known standard statistical model (with HIVAP
code \cite{Reis81,Reis85,Reis92}) is used. Then, the evaporation
residue cross sections of a series of fusion-fission reactions will
be investigated for a systematic test of the model and refining the
parameters.

\begin{center}
\textbf{II. MODIFIED WOODS-SAXON POTENTIAL AND SOME PARAMETERS OF HIVAP CODE}
\end{center}
In this section, we first introduce an empirical nucleus-nucleus
potential based on the Skyrme energy density functional within the
extended Thomas-Fermi approach. Then, the statistical model HIVAP is
briefly introduced and the influence of some key parameters is
studied. Finally, a number of calculated results are compared with
experimental data.
\begin{center}
\textbf{A.  Modified Woods-Saxon Potential and Fusion Cross Section}
\end{center}

 The nucleus-nucleus interaction potential reads as:
\begin{eqnarray}
V(R) =  V_N(R)+V_{C}(R).
\end{eqnarray}
Here, $V_N$ and $V_C$ are the nuclear and Coulomb
interactions, respectively. We take $V_C(R)=e^2 Z_1 Z_2/R$, and
the nuclear interaction $V_N$ to be of Woods-Saxon form with
five parameters determined by fitting the
entrance channel potentials obtained with the Skyrme energy density functional
within the extended Thomas-Fermi (up to second order in $\hbar$ \cite{brack}) approach proposed in \cite{Liu06}:
\begin{eqnarray}
V_N(R)=\frac{V_0}{1+\exp [(R-R_0)/a]},
\end{eqnarray}
with \cite{Dob03}
\begin{eqnarray}
V_0=u_0 [1+\kappa (I_1+I_2)]\frac{A^{1/3}_1 A^{1/3}_2}{A^{1/3}_1+
A^{1/3}_2},
\end{eqnarray}
and
\begin{eqnarray}
R_0=r_0(A^{1/3}_1+A^{1/3}_2)+c.
\end{eqnarray}
$I_1=(N_1-Z_1)/A_1$ and $I_2=(N_2-Z_2)/A_2$ in Eq.(4) are the
isospin asymmetries of projectile and target nuclei, respectively.

By varying the five free parameters $u_0$, $\kappa$, $r_0$, $c$ and
$a$ of the modified Woods-Saxon (MWS) potential, we minimize the
relative deviation between the fusion barrier height obtained with
the Skyrme energy-density functional with SkM* \cite{Bart82} force
and the barrier height of the MWS potential obtained with Eq.(2).
The corresponding optimal values of these parameters are obtained at
the minimum of the relative deviation. In this work, 66996 reactions
with $Z_1 Z_2 \le 3000$ were used to determine the parameters of the
modified Woods-Saxon potential. The obtained optimal values of the
parameters are listed in Table I. Here we also list the potential
depth parameters $u_0$ and $\kappa$ proposed in \cite{Dob03} for
comparison. In \cite{Dob03} the nuclear interaction is taken as a
Gaussian form and the potential parameters are also determined by
the Skyrme interaction SkM*. We find that the potential depth
parameters obtained with the two approaches are close to each other.

\begin{table}
\caption{ Parameters of the potential.}
\begin{tabular}{cccccc}
\hline\hline
 & $ r_0 (fm) $ & $c  (fm)$ & $u_0  (MeV)$ & $\kappa$ & $ a  (fm)$   \\ \hline
 This work & 1.27   & -1.37   & -44.16     & -0.40    & 0.75 \\
 Ref.\cite{Dob03} &   &   & -46.07 & -0.47 &  \\
 \hline\hline
\end{tabular}
\end{table}

With the modified Woods-Saxon potential together with the proposed
empirical fusion barrier distribution in \cite{Liu06}, the fusion
cross sections and the mean barrier heights of a large number of
reactions can be reproduced well \cite{Liu06,Wang06,Wang07,Tian07}.
For the reader's convenience, the empirical barrier distribution is
briefly introduced here. We assume the barrier distribution function
$D(B)$ to be a superposition of two Gaussian functions $D_1(B)$ and
$D_2(B)$,
\begin{eqnarray}
D_{1}(B)=\frac{\sqrt{\gamma}}{2\sqrt{\pi}b_{1}} \exp \left[ -\gamma \frac{%
(B-B_{1})^{2}}{(2b_{1})^{2}}\right]
\end{eqnarray}
and
\begin{eqnarray}
D_{2}(B)=\frac{1}{2\sqrt{\pi}b_{2}}\exp \left[ -\frac{%
(B-B_{2})^{2}}{(2b_{2})^{2}}\right],
\end{eqnarray}
with
\begin{eqnarray}
B_{1}=B_{c}+b_{1},
\end{eqnarray}
\begin{eqnarray}
B_{2}=B_{c}+b_{2},
\end{eqnarray}
\begin{eqnarray}
b_{1}=\frac{1}{4}(B_0-B_{c}),
\end{eqnarray}
\begin{eqnarray}
b_{2}=\frac{1}{2}(B_0-B_{c}).
\end{eqnarray}
Here $B_0$ is the barrier height from the modified Woods-Saxon
potential. The effective barrier height is $B_{c}=f B_0$ with the
reduction factor $f=0.926$. The quantity $\gamma$ in $D_1(B)$ is a
factor which empirically takes into account the structure effects
and has a value larger or equal to 0.5. For the fusion reactions
with neutron-shell open nuclei but near the $\beta$-stability line
and for the fusion reactions at energies near and above the barriers
we set $\gamma=1$. For the reactions with neutron-shell closed
nuclei or neutron-rich nuclei an empirical formula for the $\gamma$
values was proposed in \cite{Liu06}. For a more convenient
discussion, we introduce the inverse of $\gamma$ as an enhancement
factor $g=1/\gamma$. The larger the value of $g$ is, the larger the
fusion cross section at sub-barrier energies is. From the
discussions in \cite{Liu06}, we learn that for the reactions with
neutron-shell closed nuclei we have $0<g<1$ while for the reactions
with neutron-rich nuclei $1< g \le2$. With the proposed empirical
barrier distribution, and the fusion radius $R_{fus}$ and the
curvature of the barrier $\hbar\omega$ obtained with the modified
Woods-Saxon potential, the fusion excitation function (or the
capture excitation function of reactions in region III of Fig.1(a))
can be obtained (details in Refs. \cite{Liu06,Wang06})
\begin{eqnarray}
\sigma_{fus}(E_{\rm c.m.})=\min[\sigma_1(E_{\rm c.m.}),
\sigma_{\rm avr}(E_{\rm c.m.})],
\end{eqnarray}
with
\begin{eqnarray}
\sigma _{1}(E_{\rm c.m.})=\int_{0}^{ \infty }D_1(B) \; \sigma
_{fus}^{\rm Wong}(E_{\rm c.m.},B)dB,
\end{eqnarray}
and
\begin{eqnarray}
\sigma _{\rm avr}(E_{\rm c.m.})=\int_{0}^{ \infty }\left [\frac{ D_1(B)+D_2(B)}{2} \right ]\sigma
_{fus}^{\rm Wong}(E_{\rm c.m.},B)dB.
\end{eqnarray}
Where, $\sigma_{fus}^{\rm Wong}$ denotes Wong's formula
\cite{Wong73} for penetrating an one-dimensional parabolic barrier,
\begin{equation}
\sigma _{fus}^{\rm Wong}(E_{\rm c.m.},B_0)=\frac{\hbar \omega
R_{fus}^{2}}{2E_{\rm c.m.}}\ln \left( 1+\exp\left[ \frac{2\pi
}{\hbar \omega }(E_{\rm c.m.}-B_{0})\right] \right)
\end{equation}
 with the center-of-mass energy $E_{\rm c.m.}$. $B_0$, $R_{fus}$ and $\hbar\omega$
 are the barrier height, radius and curvature, respectively. The influence of angular momentum in the entrance channel has already been taken into account in Wong's formula with the assumptions that the barrier position $R_{fus}$ and the barrier curvature $\hbar\omega$ do not change with angular momentum.

\begin{figure}
\includegraphics[angle=-0,width=1.0\textwidth]{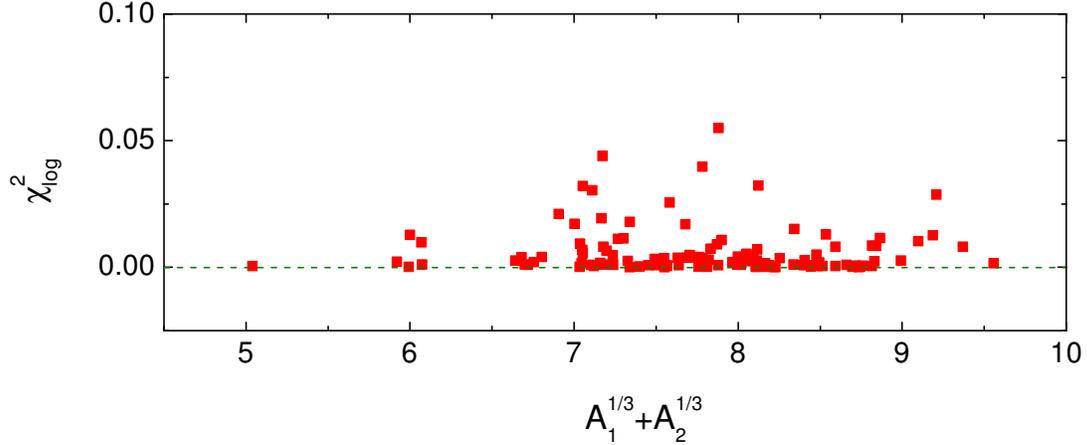}
 \caption{ (Color online) The average deviations $\allowbreak
\chi _{\log }^{2}$ for a total of 120 fusion reactions. $A_1$ and $A_2$ denote the projectile and target masses, respectively.} \label{fig2}
\end{figure}

We have calculated the fusion (capture) excitation functions of 120 fusion reactions
at energies near and above the barrier (with $g=1$) and their average deviations $\allowbreak \chi _{\log }^{2}$ from the experimental data defined as
\begin{eqnarray}
\allowbreak \chi _{\log }^{2}=\frac{1}{m}\sum_{n=1}^{m} \left[
\log (\sigma_{th}(E_n)) -\log (\sigma_{exp}(E_n))\right]^2.
\end{eqnarray}
Here $m$ denotes the number of energy-points of experimental data,
and $\sigma_{th}(E_n)$ and $\sigma_{exp}(E_n)$ are the calculated
and experimental fusion (capture) cross sections at the
center-of-mass energy $E_n$ ($E_n \ge B_{0}$), respectively. The
calculated results for  $\allowbreak \chi _{\log }^{2}$ are shown in
Fig.2. The average deviations of about $70 \%$ systems in
$\allowbreak \chi _{\log }^{2}$ are less than 0.005, with which we
can estimate the systematic error of this approach for the
description of the fusion (capture) cross sections at energies near
and above the barriers. A series of fusion reactions with $^{16}$O
bombarding on medium mass targets such as $^{144-154}$Sm are studied
with this approach, and the fusion excitation functions of these
reactions are shown in Fig.3. The energy scale has been normalized
by the mean barrier height $B_{\rm m}$ calculated with the proposed
method by setting $g=1$ \cite{Tian07}. The scattered symbols denote
the experimental data. The solid curve denote the calculated results
with $g=1$. The error bars in Fig.3(a) are estimated by $18\%$. In
Fig.3(b), we notice that nearly all of the experimental data of
sub-barrier energies are scattered in the region $0< g\le 2$ as we
defined in the proposed approach. The fusion cross sections (solid
circles) of the reactions with neutron-shell closed nuclei at
sub-barrier energies are systematically lower than the calculated
results with $g=1$ which is consistent with our discussion mentioned
previously. With $g\simeq 0$ and $g=2$ we  estimate the lower and
upper limits of the fusion (capture) cross sections at sub-barrier
energies respectively.

\begin{figure}
\includegraphics[angle=-0,width=1.0\textwidth]{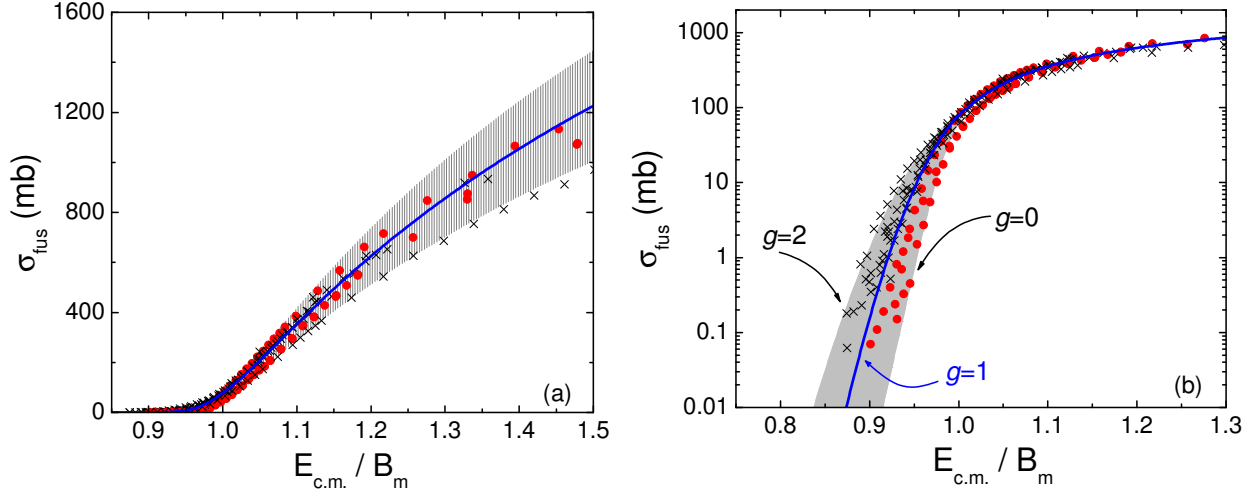}
 \caption{ (Color online) The fusion excitation functions of a series of reactions with $^{16}$O bombarding on medium mass targets. The incident energies are normalized by the mean barrier heights $B_{\rm m}$. The fusion cross sections of these reactions are shown with linear and logarithmic scale in (a) and (b), respectively. The solid circles and crosses denote the experimental data of reactions with neutron-shell closed nuclei and with neutron-shell open nuclei, respectively. The solid curve denotes the calculation result with $g=1$. The error bars in (a) are estimated with $18\%$ of the fusion cross sections. The upper and lower limits of the cross sections in (b) are obtained with $g=2$ and $g\simeq 0$, respectively.}
\end{figure}

\begin{center}
\textbf{B.  Fission Barrier and Level Density Parameter in Evaporation Calculations}
\end{center}

The calculations of the survival probabilities $W_{\rm sur}$ of the
compound nuclei were performed with the statistical evaporation code
called HIVAP which uses standard evaporation theory and takes into
account the competition of $\gamma$-ray, neutron, proton,
$\alpha$-particle emission with fission using angular-momentum and
shape-dependent two-Fermi-gas-model level density
formula\cite{Reis92}. Although it is a standard statistical model
for describing the de-excitation process, one has to reconsider some
parameters adopted for describing a wide range of fusion-fission
reactions. The sensitive parameters involved are primarily fission
barriers and level density parameters.

In the standard HIVAP code, the fission barrier at zero angular momentum is calculated by
\begin{eqnarray}
B_f=B_f^{\rm Mac}-S.
\end{eqnarray}
The macroscopic barrier $B_f^{\rm Mac}$ is usually described with a
liquid-drop model refined by Cohen and Swiatecki \cite{Co63}, Sierk
\cite{Sierk}, and Dahlinger \emph{et al.} \cite{Dah82}. The shell
correction $S$ is calculated from the difference of the experimental
mass and the liquid-drop mass, $S=M_{\rm exp}-M_{\rm LD}$. In this
code, the liquid-drop mass is calculated with the parameter set
proposed by Myers and Swiatecki in 1967 \cite{Myer67}, and the
$M_{\rm exp}$ is in fact taken from the mass table of M\"oller-Nix
\cite{Moll95} which was obtained with the finite range droplet model
and has an rms deviation of only 0.656 MeV for 2149 measured masses
of nuclei \cite{Buch05}.
\begin{figure}
\includegraphics[angle=-0,width=1.0\textwidth]{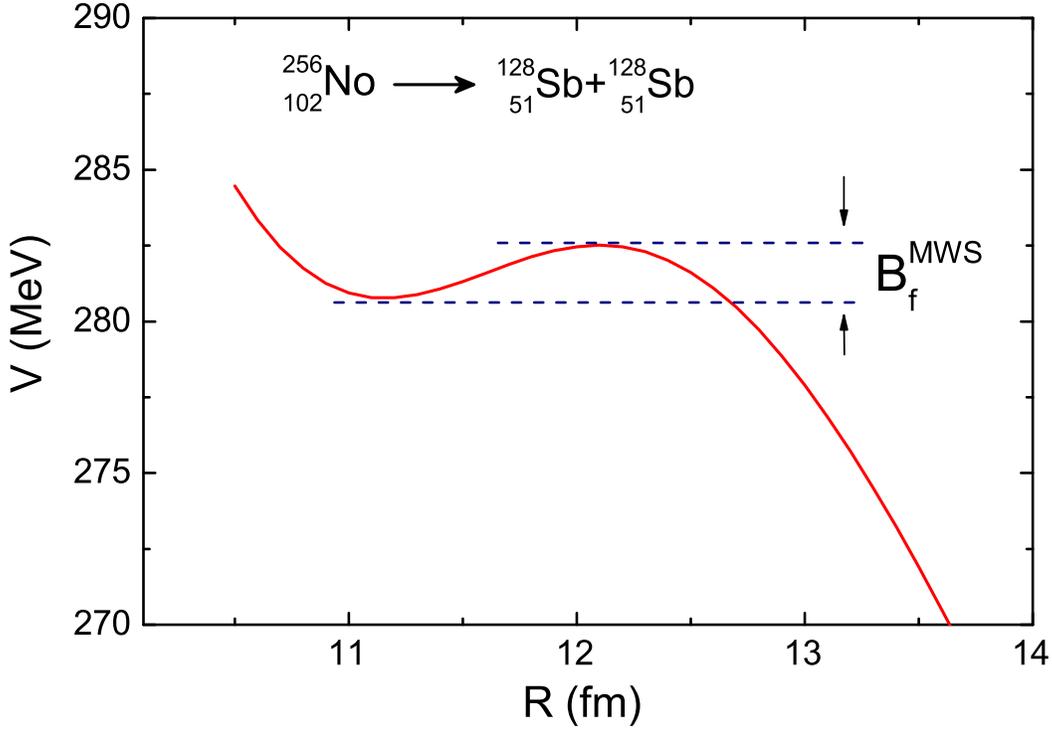}
 \caption{ (Color online) The macroscopic fission barrier $B_f^{\rm MWS}$ for $^{256}_{102}$No fissioning into two $^{128}_{\; \; 51}$Sb obtained with the modified Woods-Saxon potential. }
\end{figure}
In the present work, we calculate the macroscopic fission barriers
with the proposed modified Woods-Saxon (MWS) potential model in
which the parameters of MWS potential are obtained based on the
Skyrme energy density functional. The value of $B_f^{\rm Mac}$ is
empirically estimated by the depth of the potential pocket, as shown
as an example in Fig.4.  This figure is for the $^{256}_{102}$No
(formed in reaction $^{48}$Ca+$^{208}$Pb) fissioning into two
$^{128}_{\; \; 51}$Sb. The obtained barrier is 1.74 MeV. The
corresponding data from refs.\cite{Co63,Sierk,Dah82} are 1.44, 1.02
and 1.19 MeV, respectively. The barrier for $^{244}$Pu from our
method and from refs.\cite{Co63,Sierk,Dah82} are 4.16, 5.17, 3.95
and 4.13 MeV, respectively. The deviations between our calculated
results and the results of liquid-drop models for heavy nuclei are
in a permitting region. For medium mass nuclei, our results are in
agreement with those of refs.\cite{Beck77,Beck78} in which the
reduction of the liquid-drop barriers was discussed.

We know that the nuclear shapes during fission are more elongated
than during fusion. In this empirical approach, the neck and
elongation of the system at fission configuration can not be
described well in the sudden approximation. We concentrate on the
height of the fission barrier in this method. We will systematically
investigate 51 fusion-fission reactions with the fission barriers
obtained with four different models (MWS potential model,
Cohen-Swiatecki's\cite{Co63}, Sierk's\cite{Sierk} and
Dahlinger's\cite{Dah82} methods). The results will be discussed in
the following paragraph.

In this code, the level density is \cite{Reis81}
\begin{eqnarray}
\rho(J,E^{*})=\frac{1}{24}\left( \frac{\hbar^2}{2\theta} \right)^{3/2}(2J+1)a^{1/2}U_J^{-2}
\exp[2(a U_J)^{1/2}],
\end{eqnarray}
\begin{eqnarray}
U_J=E^{*}-E_r(J).
\end{eqnarray}
Here $E_r(J)$ is the yrast energy of either the equilibrium
configuration (light-particle and $\gamma$-emission) or the
saddle-point configuration (fission) and reads
\begin{eqnarray}
E_r(J)=J(J+1)\hbar^2/2 I,
\end{eqnarray}
in which $I$ is the moment of inertia. The level density parameter $a$ is obtained from \cite{Reis85} as
\begin{eqnarray}
a=\tilde{a}[1+f(E^{*})S/E^{*}],
\end{eqnarray}
with \cite{Ign75}
\begin{eqnarray}
f(E^{*})=1-\exp(-E^{*}/E_d)
\end{eqnarray}
with the shell damping energy $E_d$ being 18.5MeV \cite{Reis81}. In
the standard HIVAP code, the \emph{smooth}, shell-independent
level-density parameter reads
\begin{eqnarray}
\tilde{a}=0.04543\; r_a^3 A+0.1355\; r_a^2 A^{2/3} B_S + 0.1426\; r_a A^{1/3} B_K,
\end{eqnarray}
which takes into account the volume, surface and curvature
dependence of the single-particle level density at the Fermi
surface. $B_S$ and $B_K$ denote the surface and curvature factors
defined in the droplet model \cite{Myer74}. For evaporation channels
we set $B_S=B_K=1$. For the fission channel, the values of $B_S$ and
$B_K$ are tabulated as a function of the fissility parameter in
\cite{Myer74}. The ratio $\tilde{a}_f/\tilde{a}_n$ ($\tilde{a}_f$
level density parameter for fission channel, $\tilde{a}_n$ for
neutron channel) is larger than 1. It decreases towards to an unit
with the increase of the fissility parameter. The results of
$\tilde{a}_f/\tilde{a}_n$ for a series of nuclei in \cite{Reis81}
can be well reproduced. $r_a$ is the radius parameter found to be
$r_a=1.153$ fm \cite{Reis81}.

With this parametrization 51 fusion-fission reactions have been
systematically investigated with the MWS, Cohen-Swiatecki's, Sierk's
and Dahlinger's fission barriers, respectively, incorporating the
proposed approach for describing the fusion (capture) cross sections
(see Eq.(12)). Calculations of the fission and particle emission
widths with the traditional statistical theory were introduced in
\cite{Beck77}. The average deviation $\allowbreak \chi _{\log }^{2}$
(see Eq.(16)) of the evaporation (and fission) cross sections from
the experimental data for these reactions are listed in Table II. We
find that the average deviation obtained with the MWS potential is
much smaller than those obtained with the other barriers. By varying
the volume, surface and curvature coefficients in Eq.(23) and the
damping energy $E_d$, and searching for the minimum of $\allowbreak
\chi _{\log }^{2}$ with the MWS fission barriers, we find that the
values proposed by Reisdorf \cite{Reis81} (adopted in the present
work) are very close to the corresponding optimal ones. In some
references the shell damping energy was written as $E_d=k_0 A^{1/3}$
or similar forms \cite{Adam04,Feng06,Mug98}. We find that the
minimal deviation is not much improved by changing the value of the
coefficient $k_0$. Therefore, in our calculations we consequently
keep Reisdorf's coefficients, Eq.(23), that contains only one
empirically adjustable parameter $r_a$.

To determine the optimal value of $r_a$, we first study the
reasonable range of $r_a$. The level density parameter is usually
dependent on the nuclear mass number from $A/8$ to $A/12$
\cite{Wile95,Feng06,Adam04}. Fig.5 shows the level density parameter
$\tilde{a}$ as a function of nuclear mass number $A$ adopting
different values for $r_a$ ( with $B_S=B_K=1$). We estimate the
variation region of $r_a$ which ranges from about 1.075 to 1.250 fm
according to $A/12 \sim A/8$. Through a variation of $r_a$ we can
find the optimal values of $r_a$ for a certain model to describe the
fission barriers. The optimal value of $r_a$ could be different for
different fission barrier models. Through systematical investigation
of the minimal average $\allowbreak \chi _{\log }^{2}$ of the 51
fusion-fission reactions, we search for the optimal parameters set
(including the parameters of fission barrier and the $r_a$ in level
density parameter). The minimal average $\allowbreak \chi _{\log
}^{2}$ of the 51 reactions and the corresponding optimal values of
$r_a$ for the four fission barrier models are listed in Table III.
By taking the optimal values of $r_a$, the average deviations
$\allowbreak \chi _{\log }^{2}$ from the experimental data get
obviously smaller for all of these models, especially for the models
of Sierk and Dahlinger. The deviation obtained by the modified
Woods-Saxon potential is still the smallest one.

\begin{table}
\caption{ Average deviation of the evaporation (and fission) cross
sections from experimental data for 51 fusion-fission reactions with
$r_a=1.153$ fm.}
\begin{tabular}{ccccc}
\hline\hline
  model & Cohen-Swiatecki  & Sierk & Dahlinger & MWS  \\ \hline
  $\allowbreak \chi _{\log }^{2}$   & 0.2295   & 0.2177     & 0.2373    & 0.1339 \\
 \hline\hline
\end{tabular}
\end{table}

\begin{table}
\caption{ The minimal average deviation $\allowbreak \chi _{\log
}^{2}$  and the corresponding optimal value of $r_a$ adopting
different models for calculating the fission barriers.}
\begin{tabular}{ccccc}
\hline\hline
  model & Cohen-Swiatecki  & Sierk & Dahlinger & MWS  \\ \hline
  $\allowbreak \chi _{\log }^{2}$   & 0.1813   & 0.1428     & 0.1642    & 0.1086 \\
 $r_a$ & 1.106 & 1.091 & 1.095 & 1.120 \\
 \hline\hline
\end{tabular}
\end{table}

\begin{figure}
\includegraphics[angle=-0,width=0.9\textwidth]{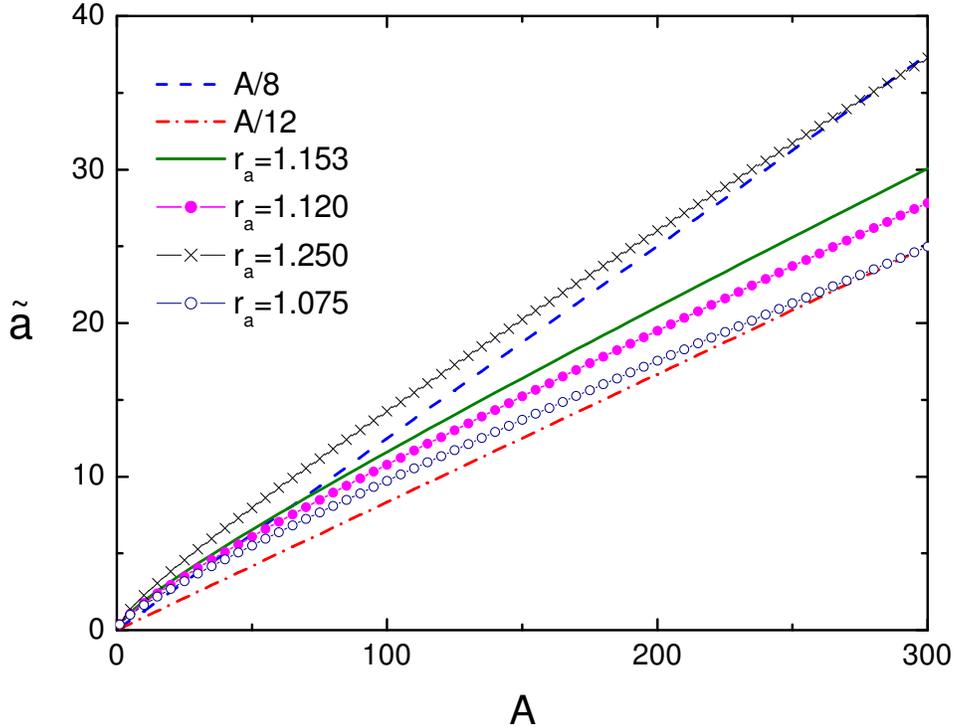}
 \caption{ (Color online) Level density parameter as a function of nuclear mass number A. }
\end{figure}

According to the formulas for the fission barrier and the level
density parameter, one learns that a reasonable calculation of the
shell correction $S$ is crucial since the shell correction plays a
role both for the fission barrier and for the level density
parameter, especially for heavy systems. Therefore, it is
interesting to compare the results with the shell corrections
obtained by different approaches for searching for the optimal
parameters set of the HIVAP code. The previous calculations
discussed are based on the shell corrections obtained with the 1967
parametrization of Mayers and Swiatecki \cite{Myer67} for $M_{\rm
LD}$. If we take the shell corrections of M\"oller-Nix based on the
1995 parametrization of the macroscopic (liquid-drop) energies of
nuclei \cite{Moll95}, the minimal deviation $\allowbreak \chi _{\log
}^{2}$ and the corresponding optimal $r_a$ are 0.1877 and 1.252 fm
with the Sierk's barrier, 0.1681 and 1.268 fm with the MWS fission
barrier, respectively. Comparing with the results listed in Table
III, one finds that using the shell corrections based on the 1967
parametrization of the liquid-drop energies $M_{\rm LD}$ of nuclei
\cite{Myer67} the fusion-fission reactions studied in this work can
be systematically better described with the present HIVAP code for
$W_{\rm sur}$ incorporating the proposed approach for
$\sigma_{fus}$. Finally, we obtain the optimal parameters set of the
HIVAP code: MWS potential model for the fission barriers, with
$r_a=1.120$ fm and together with the 1967 parametrization of the
liquid-drop energies of nuclei for the shell corrections.

\begin{center}
\textbf{C.  Comparison between the Calculated Results and the Experimental Data}
\end{center}

With the modified Woods-Saxon potential for the unified description
of the entrance channel fusion barrier and the macroscopic fission
barrier $B_f^{\rm Mac}$, with $r_a=1.120$ fm, and together with the
1967 parametrization of Mayers and Swiatecki for the shell
corrections, we obtained the deviations $\allowbreak \chi _{\log
}^{2}$ of the evaporation (and fission) cross sections from the
experimental data for the 51 fusion-fission reactions which are
shown in Fig.6. We find that $68.3\%$ reactions have values smaller
than 0.0714, with which we can estimate the upper and lower
confidence limits of the systematic errors of the HIVAP code for
$W_{\rm sur}$ (the values are  $1.85 W_{\rm sur}$ and $W_{\rm
sur}/1.85$, respectively). In the following figures, Fig.7 --
Fig.12, we present the calculated results together with the
systematic errors (the shades in the figures) of $\sigma_{fus}$ and
$W_{\rm sur}$. The experimental data are also presented for
comparison. From these figures, one finds that the experimental data
can be systematically well reproduced (within about 2 times
deviations) at energies near and above the fusion barriers.

Fig.13 shows calculated neutron evaporation residue cross sections
for heavy systems with $^{208}$Pb. Because the quasi-fission has not
been taken into account in these calculations yet, we find that the
deviations from the experimental data increase exponentially with
the increase of $Z_{\rm CN}$ (the positions of the peaks for the
evaporation residues can be roughly reproduced). This implies that
the quasi-fission plays an important role in the reactions leading
to superheavy nuclei. With the proposed approach for $\sigma_{\rm
cap}$ and $W_{\rm sur}$, the ambiguity in predicting the probability
of quasi-fission could be reduced.

\begin{figure}
\includegraphics[angle=-0,width=1.0\textwidth]{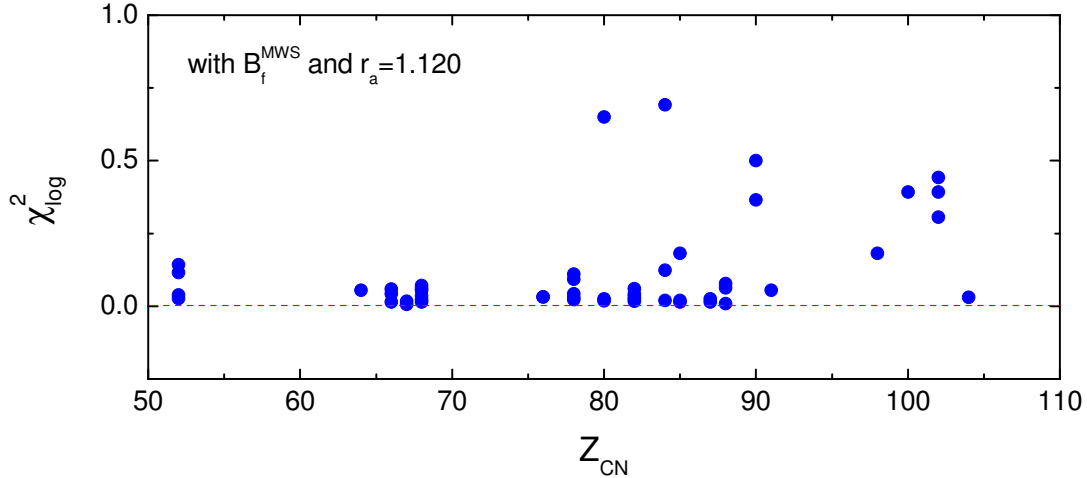}
 \caption{(Color online) Deviations $\allowbreak \chi _{\log }^{2}$ of the calculated evaporation (and fission) cross sections from the experimental data for 51 fusion-fission reactions. }
\end{figure}

\begin{figure}
\includegraphics[angle=-0,width=1.0\textwidth]{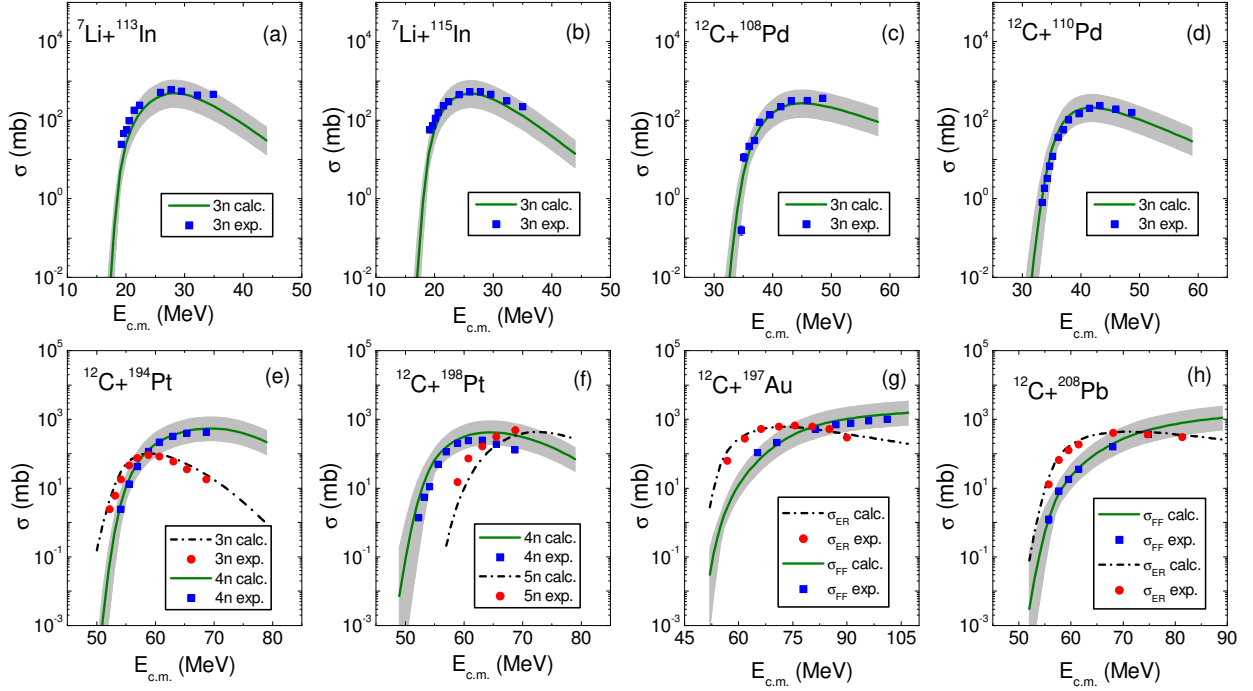}
 \caption{(Color online) The cross sections of reactions $^7$Li+$^{113,115}$In \cite{Cap96}, $^{12}$C+$^{108,110}$Pd \cite{Cap96}, $^{12}$C+$^{194,198}$Pt \cite{Shr01}, $^{12}$C+$^{197}$Au \cite{Bab88} and $^{12}$C+$^{208}$Pb \cite{San01}. $\sigma_{\rm FF}$ denotes the fission cross section. $\sigma_{\rm ER}$ denotes the evaporation residue cross section (a sum over all evaporation channels). The shades in this and the following figures denote the systematic errors of the present approach (including both the systematic errors of $\sigma_{\rm cap}$ and those of $W_{\rm sur}$), if not otherwise stated.}
\end{figure}

\begin{figure}
\includegraphics[angle=-0,width=1.0\textwidth]{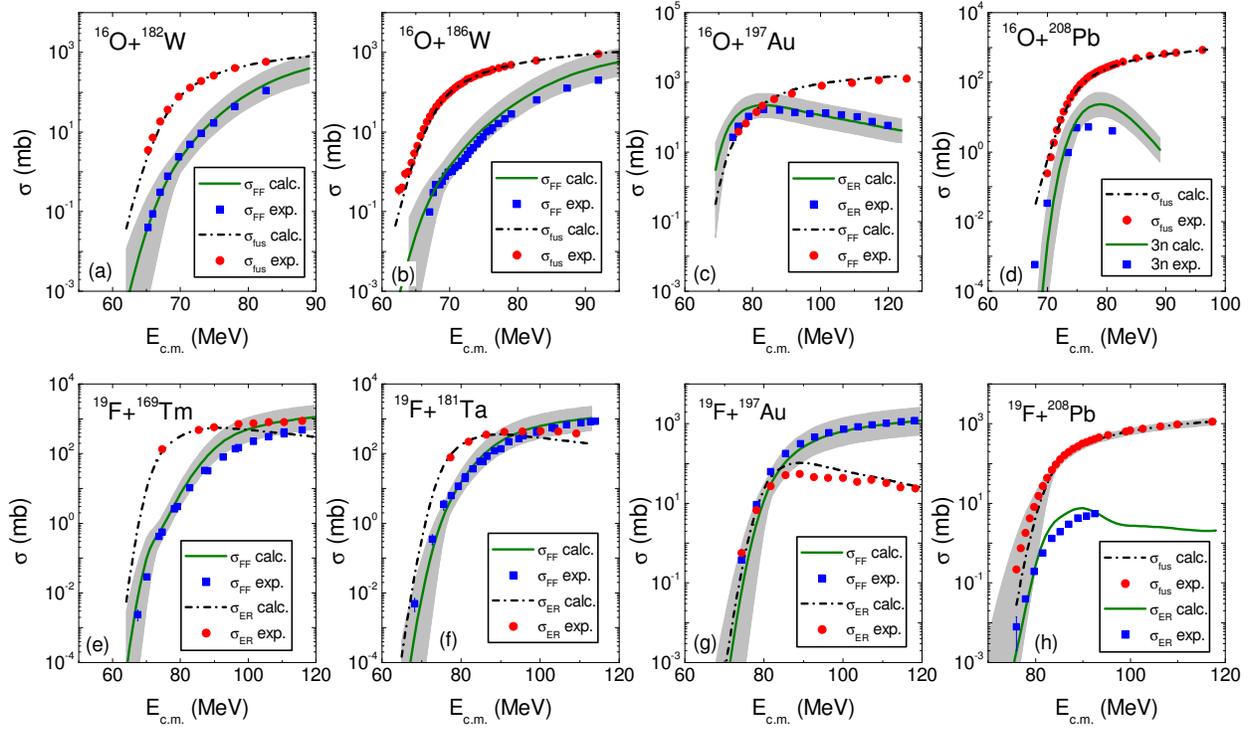}
 \caption{(Color online) The cross sections of reactions $^{16}$O+$^{182,186}$W \cite{Hin00,Lei95}, $^{16}$O+$^{197}$Au \cite{Bab88}, $^{16}$O+$^{208}$Pb \cite{Zag01}, $^{19}$F+$^{169}$Tm \cite{Cha86}, $^{19}$F+$^{181}$Ta \cite{Cha86}, $^{19}$F+$^{197}$Au \cite{Hin02} and $^{19}$F+$^{208}$Pb \cite{Hin99}. The shade in (h) denotes the systematic errors of the capture cross sections.  }
\end{figure}

\begin{figure}
\includegraphics[angle=-0,width=1.0\textwidth]{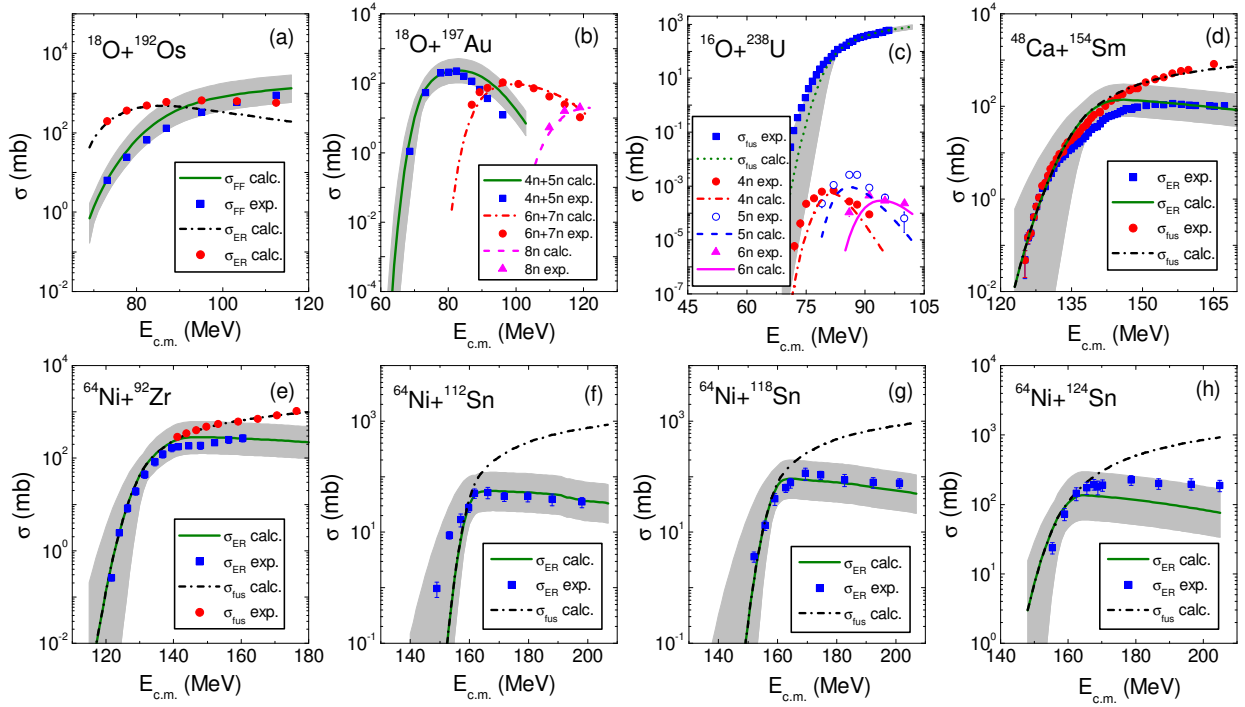}
 \caption{(Color online) The cross sections of reactions $^{18}$O+$^{192}$Os \cite{Cha86}, $^{18}$O+$^{197}$Au \cite{Cor05}, $^{48}$Ca+$^{154}$Sm \cite{Tro05}, $^{16}$O+$^{238}$U \cite{Nis04}, $^{64}$Ni+$^{92}$Zr \cite{Ste92}, $^{64}$Ni+$^{112,118,124}$Sn \cite{Liang07}. The shade in (c) denotes the systematic errors of the capture cross sections.  }
\end{figure}

\begin{figure}
\includegraphics[angle=-0,width=1.0\textwidth]{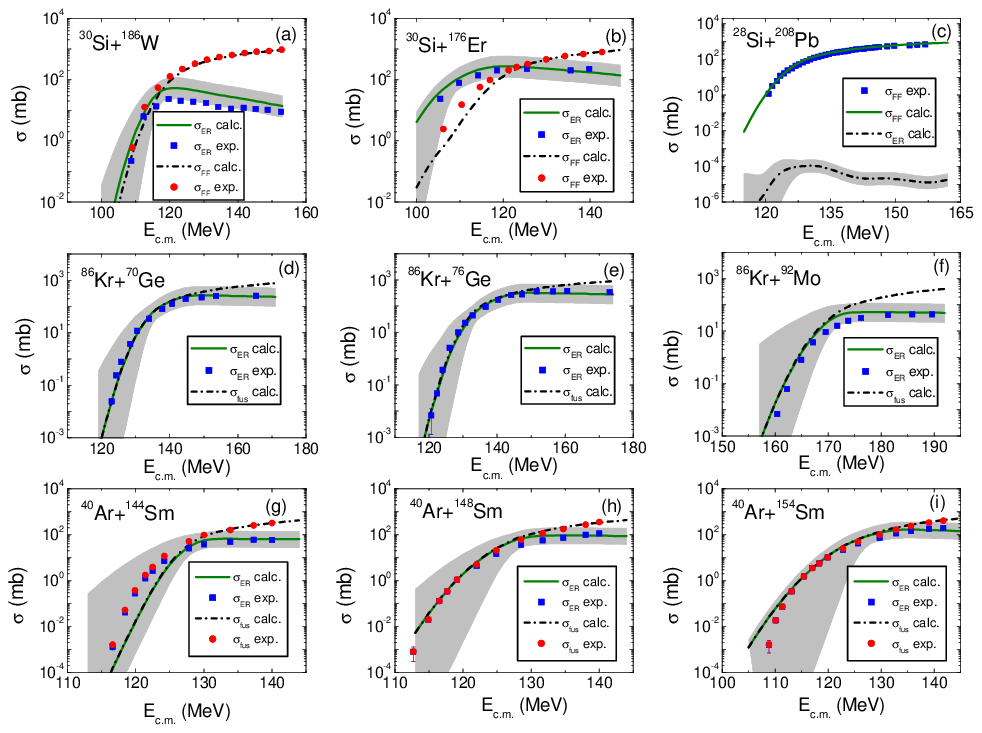}
 \caption{(Color online) The cross sections of reactions $^{30}$Si+$^{186}$W \cite{Hin02}, $^{30}$Si+$^{176}$Er \cite{Hin83}, $^{28}$Si+$^{208}$Pb \cite{Hin95}, $^{86}$Kr+$^{70,76}$Ge \cite{Reis85} and $^{86}$Kr+$^{92}$Mo \cite{Reis85}, $^{40}$Ar+$^{144,148,154}$Sm \cite{Reis85a}.   }
\end{figure}

\begin{figure}
\includegraphics[angle=-0,width=1.0\textwidth]{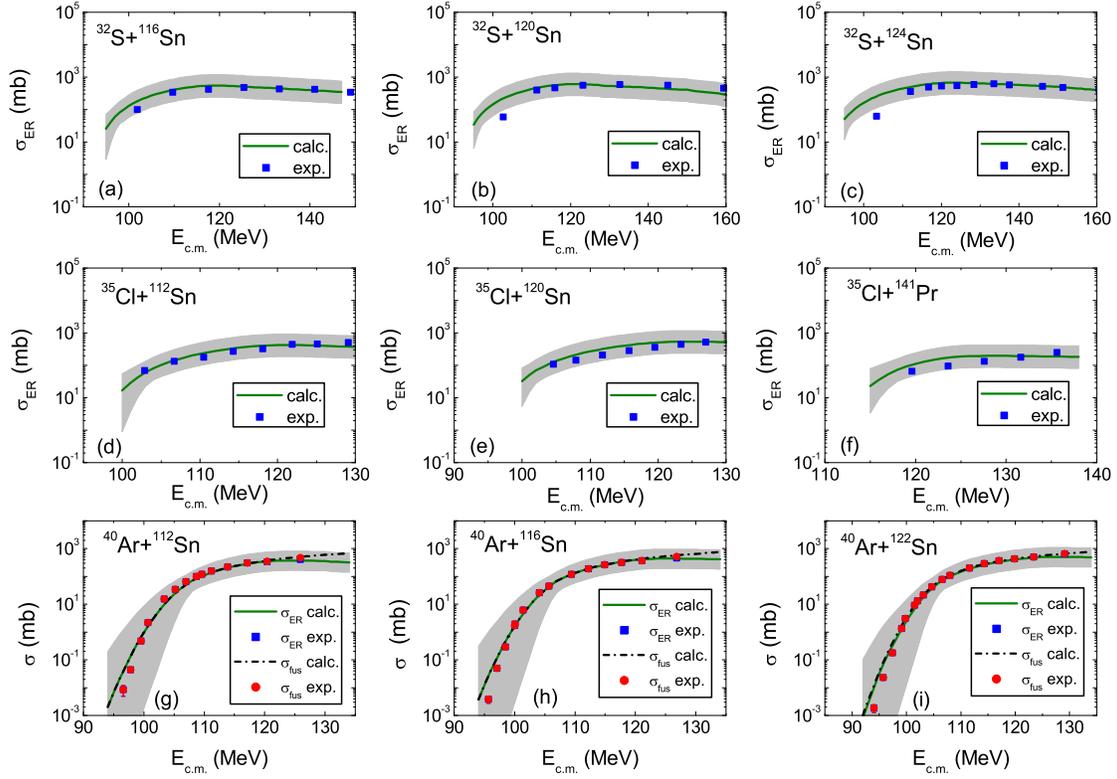}
 \caption{(Color online) The evaporation residue cross sections of reactions $^{32}$S+$^{116,120,124}$Sn \cite{Ern84}, $^{35}$Cl+$^{112,120}$Sn \cite{Dav77} and $^{35}$Cl+$^{141}$Pr \cite{Dav77}, $^{40}$Ar+$^{112,116,122}$Sn \cite{Reis85a}. }
\end{figure}

\begin{figure}
\includegraphics[angle=-0,width=1.0\textwidth]{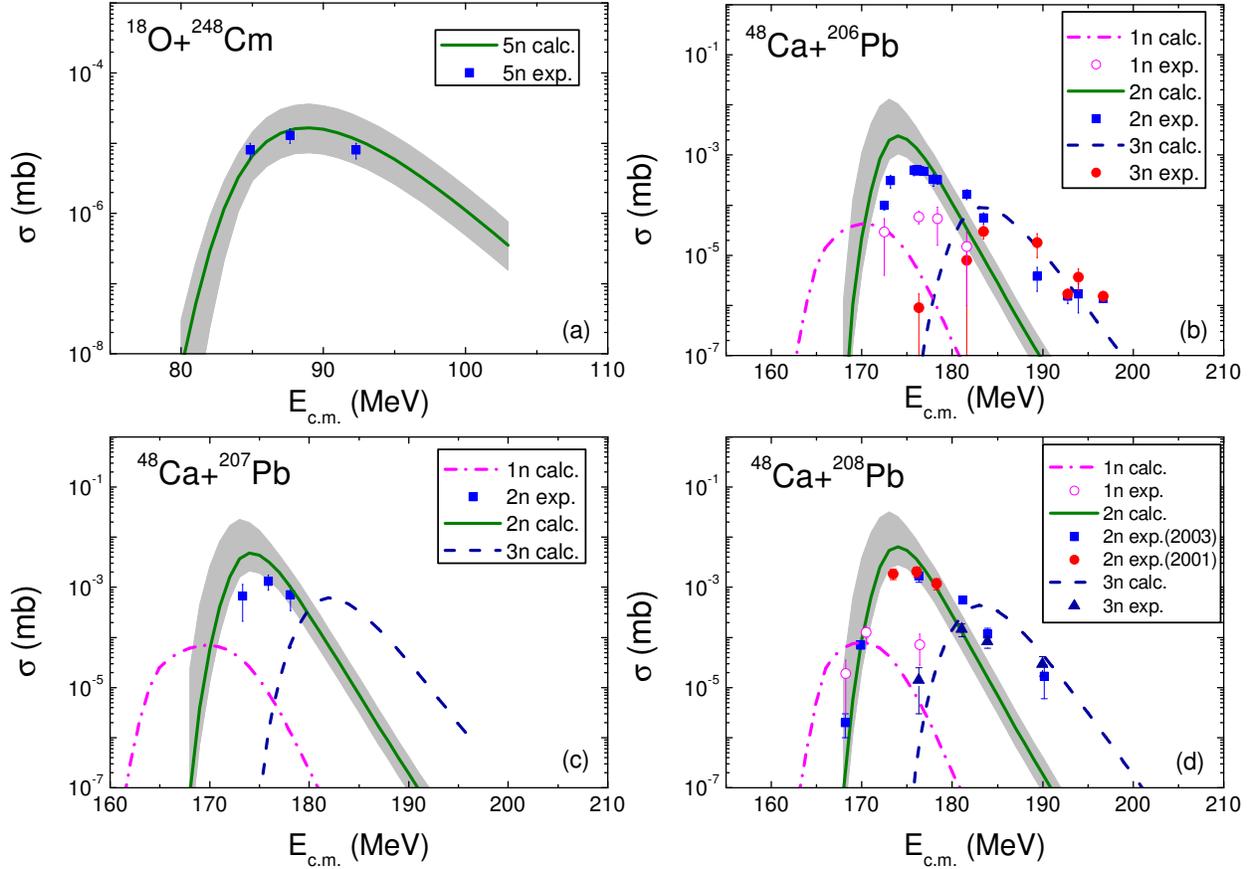}
 \caption{(Color online) The cross sections of reactions $^{18}$O+$^{248}$Cm \cite{Nag02}, $^{48}$Ca+$^{206,207,208}$Pb \cite{Oga01,Prok03}. The quasi-fission is not taken into account in the calculation.   }
\end{figure}

\begin{figure}
\includegraphics[angle=-0,width=1.0\textwidth]{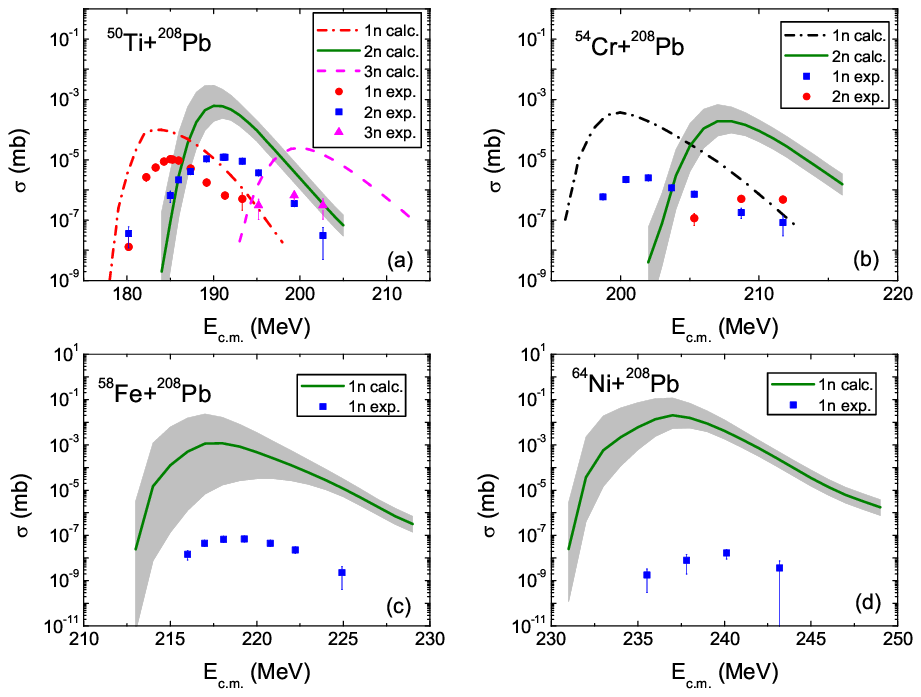}
 \caption{(Color online) The neutron evaporation residue cross sections of heavy reactions with $^{208}$Pb target \cite{Hof04,Mori04}.  The quasi-fission is not taken into account in the calculation. }
\end{figure}

\begin{center}
\textbf{III. CONCLUSION AND DISCUSSION}
\end{center}

In this work, we proposed a modified Woods-Saxon potential for a
unified description of the entrance channel fusion barrier and the
fission barrier of fusion-fission reactions which is based on the
Skyrme energy-density functional approach. With the proposed
potential for the fusion barriers, 120 heavy-ion fusion reactions
have been systematically investigated together with the barrier
penetration concept and an empirical barrier distribution. The
experimental data for the fusion cross sections $\sigma_{fus}$ can
be well reproduced and the systematic errors are $18\%$ at energies
near and above the barriers. Incorporating a statistical model HIVAP
for describing the decay of the compound nuclei, the evaporation
residue (and fission) cross sections of 51 fusion-fission reactions
have been systematically studied simultaneously to investigate and
refine some key parameters of the HIVAP code such as the fission
barrier and the level density parameter. With the optimal value of
the radius parameter $r_a=1.120$ fm of the level density parameter,
and with the fission barriers obtained by the proposed modified
Woods-Saxon potential, the experimental data can be systematically
reproduced reasonably well. The upper and lower confidence limits of
the systematic errors of the calculated survival probabilities
$W_{\rm sur}$ with the HIVAP code are $1.85 W_{\rm sur}$ and $W_{\rm
sur}/1.85$, respectively. The influence of the shell corrections on
the calculated results has been explored. The 1967 parametrization
of Mayers and Swiatecki \cite{Myer67} for the macroscopic
(liquid-drop) energies of nuclei gives better results in the case of
these 51 reactions. For the systems leading to superheavy nuclei,
the influence of quasi-fission increases rapidly with increasing the
compound-nuclear charge number $Z_{\rm CN}$. With the individual
investigation of $\sigma_{\rm cap}$ and $W_{\rm sur}$, the ambiguity
of the prediction of the evaporation cross sections could be
reduced, which is helpful in testing models for the formation
probability $P_{\rm CN}$ of compound nuclei.

In the present work, the estimated systematic errors based on the
enhancement factor $0<g\le 2$ for the capture cross sections at
sub-barrier energies are still large, especially for heavy systems.
A precise prediction of the enhancement factor $g$ and a reduction
of the corresponding systematic errors are still required,
especially for the "cold fusion" in which the suitable incident
energies for producing evaporation residues are near or lower than
the average fusion barrier. For "hot fusion" systems, the suitable
incident energies could be higher than the average fusion barrier,
and thus the influence of $g$ decreases since the capture cross
sections are not very sensitive to the enhancement factor $g$ at
energies above the barrier. In addition, the influence of asymmetric
fission and the time-dependent fission width \cite{Jura05} have not
been taken into account yet. It is known that the nuclear
dissipation influences the saddle-to-scission time and thus
influences the competition between fission and particle evaporation.
These effects are very important in fission dynamics but they are
beyond the scope of this work. Work on these aspects is in progress.

\newpage

\begin{center}
\textbf{ACKNOWLEDGEMENTS}
\end{center}

One of the authors (N. Wang) thanks for the Alexander von Humboldt Foundation for support.

\end{document}